\documentclass[%
letterpaper,
final,
tightenlines,
raggedbottom,
showpacs,
nofootinbib,
twoside,
twocolumn,
pdfstartview=FitH,
10pt,
amsmath,amssymb,amsfonts,
aps,
pra,
floatfix,
]{revtex4-1}

\usepackage[T1]{fontenc} % for copiable characters with accents over them

\input{glyphtounicode}
\allowdisplaybreaks

\usepackage[sort&compress]{natbib}
\usepackage[
	bookmarks,
	breaklinks,
	colorlinks,
	allcolors=blue,
	urlcolor=blue,
	linktocpage,
	hyperfootnotes,
	bookmarkstype=toc,
	bookmarksopen,
	bookmarksopenlevel=1
]{hyperref}
\hypersetup{pdfpagemode=UseNone}

\usepackage{wasysym}  % gives lovely straight integral sign

\usepackage{textcomp}
\usepackage{graphicx}% Include figure files
\usepackage{dcolumn}% Align table columns on decimal point
\usepackage{bm}% bold math
\usepackage[all]{hypcap}

\usepackage{xcolor}
\definecolor{grey}{rgb}{0.5,0.5,0.5}
\definecolor{lightgrey}{rgb}{0.9,0.9,0.9}

\makeatletter
\renewcommand*{\p@section}{}
\renewcommand*{\p@subsection}{}
\renewcommand*{\p@subsubsection}{}
\makeatother

\begin{document}

\title{Ultracold atom interferometry with pulses of variable duration}


\author{Valentin Ivannikov}
%\thanks{}
\affiliation{Centre for Quantum and Optical Science, Swinburne University of Technology, Melbourne, Australia}
\email{valentin@ifsc.usp.br}
\altaffiliation[Present address: ]{Instituto de F\'isica de S\~ao Carlos, Universidade de S\~ao Paulo, Avenida Trabalhador S\~ao-Carlense, 400, S\~ao Carlos, S\~ao Paulo, CES 13566-590, Brazil.}

\date{16 March 2017}

\begin{abstract}
We offer interferometry models for thermal ensembles with one-body losses and the phenomenological inclusion of perturbations covering most of the thermal atom experiments. A possible extension to the many-body case is briefly discussed. The Ramsey pulses are assumed to have variable durations and the detuning during the pulses is distinguished from the detuning during evolution. Consequently, the pulses are not restricted to resonant operation and give more flexibility to optimize the interferometer to particular experimental conditions. On this basis another model is devised in which the contrast loss due to the unequal one-body population decays is cancelled by the application of a non-standard splitting pulse. For the importance of its practical implications, an analogous spin-echo model is also provided. The developed models are suitable for the analysis of atomic clocks and a broad range of sensing applications, they are particularly useful for trapped-atom interferometers.
\end{abstract}

%\keywords{}

\maketitle

\section{Introduction}

Most accurate experimental methods have been based on interferometers, first invented for the measurement of the velocity of light and gradually extended to frequency standards and metrology \cite{Fizeau1851,Michelson1881,Rayleigh1881,Michelson1998,Rabi1938,Rabi1939,Ramsey1949,Ramsey1990,Cronin2009}. In recent years the interest in precision interferometry has been growing in the context of ultracold atomic systems \cite{Doring2009,Gross2010,Hinkley2013b}, where atom chips that allow unparalleled control over atomic ensembles have become particularly promising \cite{Riedel2010a,Egorov2011}.

In metrology it becomes a burden to interpret how inadvertently detuned pulses quantitatively influence results. Our analytical models address most common cases. Imperfections are often compensated with sophisticated techniques \cite{Wimperis1989,Wimperis1994,Uhrig2007,Levitt2007} that may become an absolute necessity: The medium may distort the pulses in an uncontrolled way or the pulse sources may suffer from imperfections and lead to measurement inaccuracies. In the high-precision applications one employs the longest possible times, limited by the saddle point of the Allan variance graph \cite{Allan1966} showing that longer integration times will gain no greater accuracy. As a consequence, the analysis should include particle losses and cold atomic collisions since they cause ensemble dephasings. In dense thermal clouds two-body processes may become a limiting factor; thus, they should also be included \cite{Ivannikov2014}.

In this contribution we offer a set of analytical models of various Ramsey-type interferometers \cite{Ramsey1950,Sullivan2001} with one-body losses and proper accounting for off-resonant coupling and ensemble dephasing. The models cover most of the experiments with thermal atoms and are extendable via the inclusion of perturbations. The extension to the case of two-body losses can be readily implemented \cite{Ivannikov2014}. A peculiar feature of the presented models is that the coupling field detuning from the atomic resonance is distinguished from the energy level shifts during free evolution, and, since the effect of the off-resonant interrogation is often significant, the detuning is assumed non-zero. The models are linked to the Bloch vector formalism to introduce the measurables and explain the underlying processes. Second, a more generic Ramsey-model is developed where both pulses have variable durations. Then a method to maximize the visibility by applying a non-$\pi/2$ splitter pulse is devised. It is based on the fact that the initialization of the two states having unequal population decays may, by the end of the evolution, result in the population equalization highly desirable for result interpretation. Ramsey spectra with Rabi pedestals are given analytically for the one-body models. In \hyperref[sec:measurables]{Appendix~\ref*{sec:measurables}} we extend the Ramsey formalism to the off-resonant spin echo with one-body losses.

\section{Interferometry models}
\label{sec:onebodyloss}

\begin{figure}[t]
\includegraphics[width=0.48\textwidth]{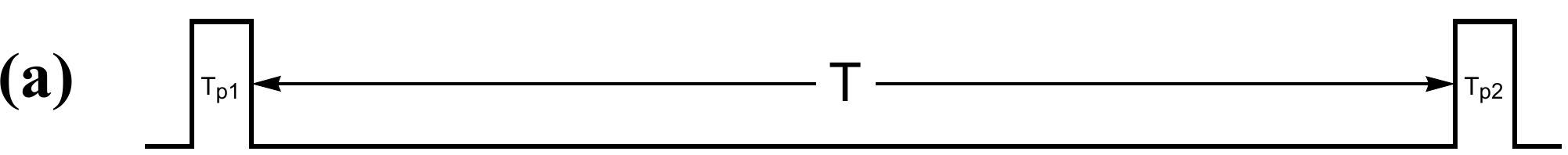} \\[2mm]
\includegraphics[width=0.48\textwidth]{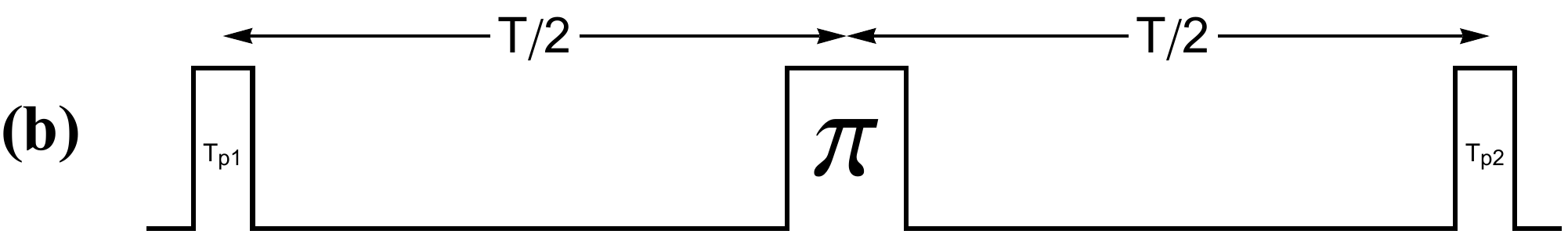}
\caption[Ramsey interferometry with variable-duration pulses]{Ramsey interferometry with variable-duration pulses.\\
\textbf{(a)} Pulses are generally non-$\pi/2$ with durations $T_{p1}$ and $T_{p2}$.\\
\textbf{(b)} Spin-echo sequence with variable-duration pulses.}
\label{fig:ramseyvar}
\end{figure}

\subsection{Definitions}
\label{sec:timeramsey}

\begin{figure*}[pt]
\includegraphics[width=0.96\textwidth]{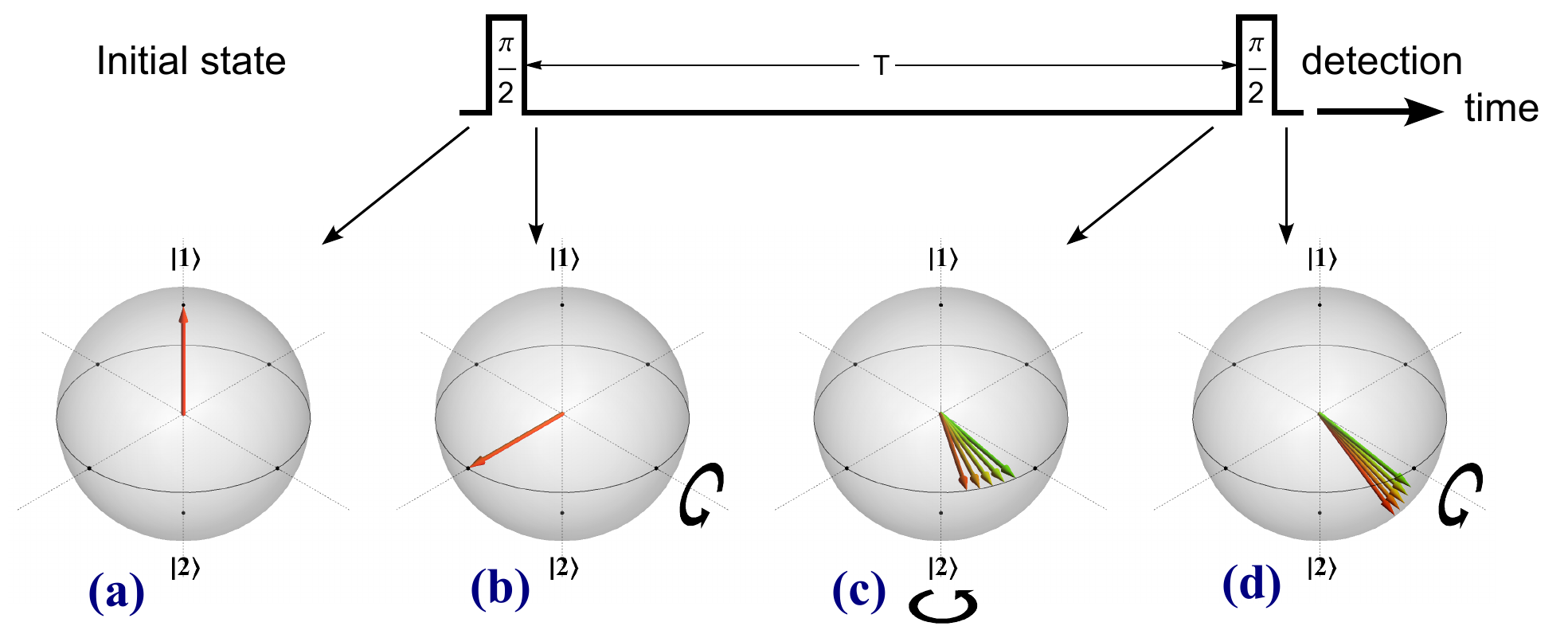}
\caption[Bloch vector evolution in Ramsey interferometry]{(Color available online) Ramsey interferometry in the Bloch vector representation. The fans of vectors illustrate phase diffusion. The black circular arrows show the rotation by the effective torque acted on the vector. A fringe in $P_z$ is produced by varying $T$. In the phase-Ramsey method the fringe is produced at a fixed $T$ by varying the phase of the second $\pi/2$-pulse.\\
\textbf{(a)} Initially the atoms are in state $\left|1\right\rangle$, state $\left|2\right\rangle$ is unpopulated.\\
\textbf{(b)} After the first $\pi/2$-pulse the effective torque brings the vector to the phase-plane ($P_z=0$).\\
\textbf{(c)} Without interrogation the system relaxes, undergoing population loss and phase destruction, resulting in phase diffusion.\\
\textbf{(d)} The second $\pi/2$-pulse applies an effective torque projecting the vectors onto the axis of population difference.}
\label{fig:blochramsey}
\end{figure*}

The ensemble of two-level atoms is initially prepared in the state $\left|1\right\rangle$. A short $\pi/2$ Rabi pulse (Fig.~\ref{fig:ramseyvar}(a)) equates the populations placing the ensemble (pseudo-)spin to the equatorial plane of the Bloch sphere where it can only evolve in phase (Fig.~\ref{fig:blochramsey}). We assume the pulses are instantaneous and neglect losses by satisfying $\Omega \gg \Delta$ and $1/\gamma_m \gg d$ with $m \in \{1,2\}$ for the resonant Rabi frequency $\Omega$, detuning from the atomic resonance $\Delta = \omega_{\text{atom}} - \omega_{\text{light}}$, population loss rates $\gamma_m$ of the states $\left|m\right\rangle$, and the pulse duration $d$. The pulse areas equal $\pi/2$ at any $\Delta$ and have duration $\pi\left/\left(2\Omega_R\right)\right.$, with the generalized Rabi frequency $\Omega_R = \sqrt{\Delta^2 + \Omega^2}$, to preserve the $\pi/2$-behavior away from resonance. It is distinct from the usual duration $\pi\left/\left(2\Omega\right)\right.$ yielding a non-$\pi/2$ pulse at $\Delta\neq 0$. The spectra of such systems differ as shown in the forthcoming discussion.

After the first $\pi/2$-pulse, the system evolves for a time $T$. The phase difference between the two states starts growing. Before the second $\pi/2$-pulse arrives, the phase is diffused due to the ensemble-related dephasing, trap-induced dephasing, and the driving frequency instability. The second $\pi/2$-pulse rotates the Bloch vector to accomplish projective detection. It brings the imprinted phase to the axis of the normalized population difference $P_z$. Locally dephased parts of the ensemble result in a blurred distribution of $P_z$, the width of which expresses the detection limit.

The measurables are expressed in terms of the atom numbers $N_{n}$ and density matrix elements $\rho_{nn}$ with the state index $n\in \{1,2\}$, $N = N_1 + N_2$; $P_z = P_1 - P_2$:
\begin{equation}
P_1 = \frac{N_1}{N} = \frac{\rho_{11}}{\rho_{11} + \rho_{22}}, \quad P_2 = \frac{N_2}{N} = \frac{\rho_{22}}{\rho_{11} + \rho_{22}}.
\label{eq:measurables}
\end{equation}

The Bloch vector $\mathbf{B}$ is employed in Fig.~\ref{fig:blochramsey} to articulate the processes in the effective two-level system. The Liouville--von Neumann equation for the resonant lossless case reads $\partial\mathbf{B}/\partial t = \mathbf{\Omega}\times\mathbf{B}$, where $\mathbf{B}=(B_x, B_y, B_z)^{\top} = \left(\rho_{21} + \rho_{12}, \operatorname{Im}\!\left\{\rho_{21} - \rho_{12}\right\}, \rho_{11} - \rho_{22}\right)^{\top}$ is the pseudo-spin vector, and $\mathbf{\Omega} = (-\Omega, 0, \Delta)^{\top}$ acts on $\mathbf{B}$ as an effective torque. During free evolution, $\varphi$ accumulates detuning and miscellaneous perturbations, e.g., collisional level shifts, radiation shifts, etc, in general taking the form of a sum $\varphi = \Delta + \Delta_{\text{collisions}} + \Delta_{\text{radiation}} + \ldots$. Separation of pulse $\Delta$ and level shifts during evolution $\varphi$ enables the model to sense perturbations. We shall refer to $\varphi$ as the Ramsey dephasing rate measured in $\text{rad}/\text{s}$. The presented models are parametrized by the Ramsey evolution time $T$, the cumulative Ramsey dephasing rate $\varphi$ during free evolution, and the phenomenological dephasing rate $\gamma_d$.

\subsection{Master equation with one-body losses}
\label{sec:liouveq}

Particle loss that causes dephasing, and the pure dephasing that only occurs between the states and is not associated with population loss, can be included in the Liouville--von Neumann equation. It is then written for a two-level system as
\begin{gather}
\frac{\partial\rho}{\partial t} = \frac{1}{i\hbar}\left[\mathbf{H}, \rho \right] - \frac{1}{2}\left\{\Gamma, \rho\right\} - \frac{1}{2} \Xi \circ \! \rho,
\label{eq:master}
\end{gather}
where $\Gamma$ is the loss operator that sets up $\gamma_1$, the population loss rate of state $\left|1\right\rangle$, and $\gamma_2$, the population loss rate of state $\left|2\right\rangle$. $\rho$ is the density operator, $[\bullet,\bullet]$ and $\{\bullet,\bullet\}$ are commutator and anticommutator brackets, respectively, and $\mathbf{H}$ is the effective Hamiltonian of a spin-$\frac{1}{2}$ system; here we shall consider $\mathbf{H}$ in the rotating wave approximation and interaction picture. The loss operator is defined as a matrix:
\begin{equation}
\Gamma = \begin{bmatrix}
\gamma_1 & 0 \\
0 & \gamma_2
\end{bmatrix}, \quad
\Xi = \begin{bmatrix}
0 & \gamma_{12} \\
\gamma_{21} & 0
\end{bmatrix}.
\label{eq:lossmatrix}
\end{equation}
The Hadamard product allows us to conveniently introduce the off-diagonal phase relaxation rates $\gamma_d$ in the pure dephasing operator $\Xi$ as a separate summand $\Xi \circ \! \rho$ of Eq.~(\ref{eq:master}) where $\Xi$ takes the form of a matrix with equal pure dephasing rates: $\gamma_{12} = \gamma_{21} = \gamma_d$. Eq.~(\ref{eq:master}) can be written explicitly as the following differential equations\footnote{$\xi$ from these equations is erroneously typed as $\Delta$ in Ref.~\cite{Ivannikov2013thesis}.}:%
\begin{equation}
  \begin{aligned}
\frac{\partial\rho_{11}}{\partial t} &= -\gamma_1 \rho_{11} - \frac{i}{2} \Omega \left(\rho_{21}-\rho_{12}\right), \\
\frac{\partial\rho_{22}}{\partial t} &= -\gamma_2 \rho_{22} +\frac{i}{2} \Omega \left(\rho_{21}-\rho_{12}\right), \\
\frac{\partial\rho_{12}}{\partial t} &= -\gamma_3 \rho_{12} + \frac{i}{2}\Omega\left( \rho_{11}-\rho_{22}\right) + i \xi \rho_{12}, \\
\frac{\partial\rho_{21}}{\partial t} &= -\gamma_3 \rho_{21} - \frac{i}{2}\Omega \left(\rho_{11}- \rho_{22}\right) - i \xi \rho_{21},
  \end{aligned}
\label{eq:masterexplicit}%
\end{equation}
where $\xi$ is included in $\mathbf{H}$ and the pure dephasing rate $\gamma_d$ in relaxation constant $\gamma_3 = \frac{1}{2}\left.\left(\gamma_1 + \gamma_2 + \gamma_d\right)\right.$. We shall distinguish two regimes: during coupling pulses ($\xi=\Delta$) and during free evolution ($\xi=\varphi$).

The Liouville--von Neumann equation [Eq.~(\ref{eq:master})] is solved with non-zero loss terms included with the assumption of lossless interrogation pulses. The solution of Eq.~(\ref{eq:master}) for the off-resonant Ramsey sequence with non-negligible losses during evolution are
\begin{subequations}
\begin{align}
\!\!\!\!\rho_{11} & = \frac {1} {4\Omega_R^4} \left(\Omega^4 e^{-\gamma_2 T} + \left(\Delta^2 + \Omega_R^2\right)^2 e^{-\gamma_1 T} - k_1 \right), \\
\!\!\!\!\rho_{22} & = \frac {1} {4\Omega_R^4} \left(\Omega^2 \left(\Delta^2 + \Omega_R^2 \right) \left(e^{-\gamma_1 T} + e^{-\gamma_2 T}\right) + k_1 \right)\!, \\
\!\!\!\!P_z & = \frac{\Delta^2\left(k_2-k_3\right)-k_1 e^{\left(\gamma_1+\gamma_2\right)T}}{\Omega_R^2\left(k_2+k_3\right)},
\end{align}
\label{eq:ramseyFormulae4}%
\end{subequations}
with the following auxiliary definitions:
\begin{equation}
\begin{aligned}
k_1 & = 2 \Omega^2 e^{-\gamma_3 T} (\Omega^2 \cos(\varphi T) - 2 \Delta \Omega_R \sin(\varphi T)), \\
k_2 & = \left(\Delta^2 + \Omega_R^2 \right) e^{\gamma_2 T}, \\
k_3 & = \Omega^2 e^{\gamma_1 T}.
\end{aligned}
\label{eq:ramseyFormulae4consts}
\end{equation}

It is often the case that the interrogation is resonant with the atomic transition. Then Eqs.~(\ref{eq:ramseyFormulae4}) simplify as%
\begin{subequations}%
\begin{align}%
\rho_{11} & = \frac{1}{4} \left(e^{-\gamma_1 T}+e^{-\gamma_2 T}-2e^{-\gamma_3 T} \cos(\varphi T)\right), \\
\rho_{22} & = \frac{1}{4} \left(e^{-\gamma_1 T}+e^{-\gamma_2 T}+2e^{-\gamma_3 T} \cos(\varphi T)\right), \\
P_z & = -e^{-\frac{1}{2}\gamma_d T}\operatorname{sech}\left(\frac{\gamma_1-\gamma_2}{2}T\right)\cos(\varphi T).
\end{align}%
\label{eq:ramseyFormulae4resonant}%
\end{subequations}%
%Equation~(\ref{eq:ramseyFormulae4resonant}c) owes its simplicity to the inherent symmetry of $P_z$ due to normalization to $N(t)$.

\subsection{Interferometry with variable-duration pulses}
\label{sec:unequalramsey}
% 12-10-01. Thesis calculations 6 (Ramsey) v4 - non-pi2 pulses.nb

As a useful extension to the Ramsey technique we present solutions for the interferometry with variable durations of the splitting and detecting pulses.

% % 12-10-01. Thesis calculations 6 (Ramsey) v4 - non-pi2 pulses.nb
\begin{figure*}[t]
\includegraphics[width=0.96\textwidth]{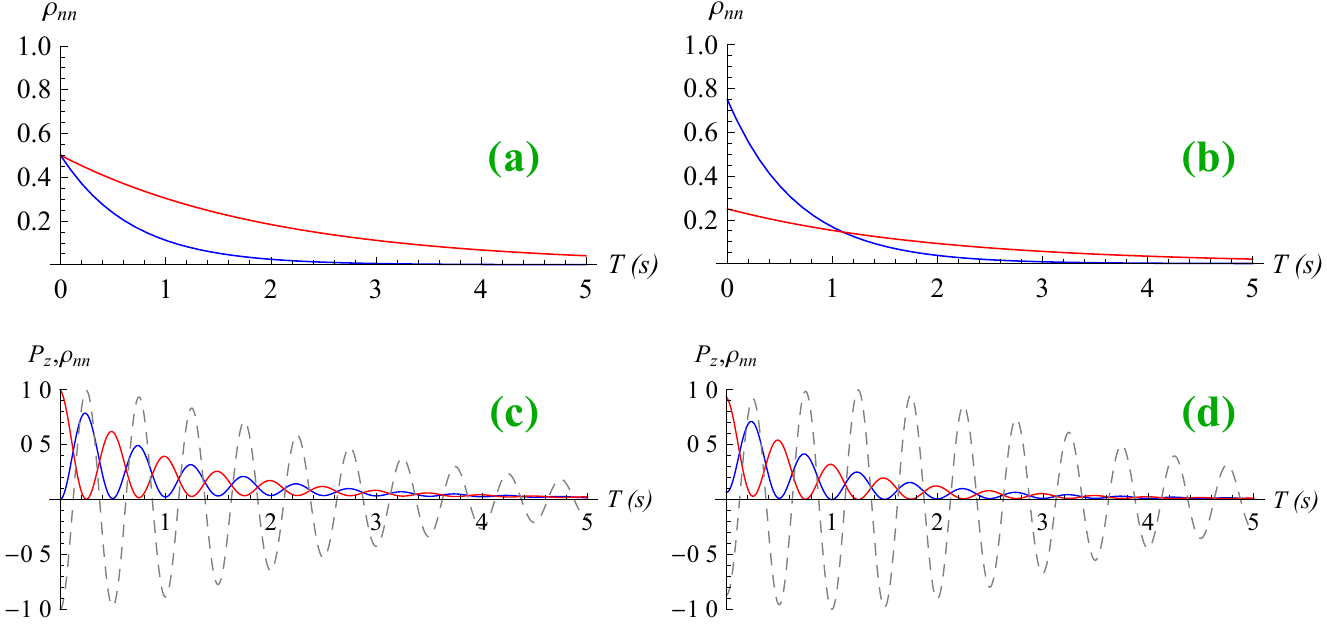}
\caption[Equalizing the state $\left|1\right\rangle$ and $\left|2\right\rangle$ instantaneous population decays]{(Color available online) Equalizing the state $\left|1\right\rangle$ and $\left|2\right\rangle$ instantaneous population decays by a variable splitter pulse to enhance $P_z$. The model is evaluated for the following parameters: $\Delta=0$~rad~s$^{-1}$, $T_{p2} = \pi/(2\Omega_R)$, $\gamma_1 = 1.5$~s$^{-1}$, $\gamma_2 = 0.5$~s$^{-1}$, $\gamma_d = 0$~s$^{-1}$, $\varphi = 2\pi\times 2$~rad, and $\Omega = 2\pi\times 1$~rad~s$^{-1}$. $\rho_{11}$ (blue) and $\rho_{22}$ (red) are from Eqs.~(\ref{eq:lossprob}), $P_z$ (dashed) from Eq.~(\ref{eq:ramseyvar1}).\\
\textbf{(a)} $T_{p1} = \pi/2$: equally split populations $\rho_{nn}$ at the end of free evolution, at $T$;\\
\textbf{(b)} $T_{p1} = \pi/3$: unequal splitting to produce a crossing of the population decays before the arrival of the second pulse;\\
\textbf{(c)} $T_{p1} = \pi/2$: after the complete sequence, $P_z$ has a monotonic $\operatorname{sech}\!\left[(\gamma_1-\gamma_2)T/2\right]$ envelope at $\gamma_1 \neq \gamma_2$ according to Eq.~(\ref{eq:ramseyFormulae4resonant}c);\\
\textbf{(d)} $T_{p1} = \pi/3$: after the complete sequence a peak of visibility is observed at $T = T_\text{optimal} \approx 1.1$~s as expected from Eq.~(\ref{eq:optimalt1}).}
\label{fig:asymcompens}
\end{figure*}

The Liouville--von Neumann Eq.~(\ref{eq:master}) is solved with the following assumptions: the detuning of the interrogating field $\Delta$ is arbitrary, no losses during the interrogating pulses, no coupling during free evolution (i.e., $\Omega = 0$). Then $\rho_{11}$ and $\rho_{22}$ at the interferometer output are
\begin{equation}
\begin{aligned}
\!\!\!\rho_{11} \,& 4 \Omega_R^4 e^{(\gamma_1+\gamma_2)T} = 4\Omega^4 e^{\gamma_1 T}\sin^2\!\!\left[\frac{\Omega_R}{2} T_{p1} \right] \sin^2\!\!\left[\frac{\Omega_R}{2} T_{p2} \right] \\
		& + k_3 k_4 e^{-\gamma_1 T} - 2 \Omega^2 e^{\frac{1}{2}(\gamma_1+\gamma_2-\gamma_d)T} (k_1 - k_2), \\
\!\!\!\rho_{22} \, & 4 \frac{\Omega_R^4}{\Omega^2} e^{(\gamma_1+\gamma_2)T} = - k_4 \cos(\Omega_R T_{p1}) - k_3 \cos(\Omega_R T_{p2}) \\
		& +k_3 + k_4 + 2e^{\frac{1}{2}(\gamma_1+\gamma_2-\gamma_d)T} (k_1 - k_2),
\end{aligned}
\label{eq:ramseyvarpops}%
\end{equation}
with auxiliary definitions for the sake of compactness:
\begin{align}
k_1 &= \cos(\varphi T)\bigg[ \Omega_R^2 \sin\!\left(\Omega_R T_{p1} \right) \sin\!\left(\Omega_R T_{p2} \right) \bigg. \nonumber\\
	& \quad\bigg. - 4 \Delta^2 \sin^2\!\left(\frac{\Omega_R}{2} T_{p1} \right) \sin^2\!\left(\frac{\Omega_R}{2} T_{p2} \right)\bigg], \nonumber\\
k_2 &= \sin\!\left(\varphi T\right) \Omega_R \Delta \left[ \sin\!\left(\Omega_R T_{p1}\right) + \sin\!\left(\Omega_R T_{p2}\right) \right. \nonumber\\
	& \quad\left. - \sin\!\left(\Omega_R T_{p1} + \Omega_R T_{p2}\right) \right], \\
k_3 &= e^{\gamma_2 T} \left(\Delta^2 + \Omega_R^2 + \Omega^2 \cos\!\left(\Omega_R T_{p1} \right) \right), \nonumber\\
k_4 &= e^{\gamma_1 T} \left(\Delta^2 + \Omega_R^2 + \Omega^2 \cos\!\left(\Omega_R T_{p2} \right) \right), \nonumber \\
k_5 &= e^{\gamma_1 T} \Omega^2 \left(1 - \cos\!\left(\Omega_R T_{p1}\right) \right). \nonumber
\label{eq:ramseyvarconsts}
\end{align}
The normalized population difference is then
\begin{equation}
\begin{aligned}
P_z &= \frac{\Omega_R^{-2} }{k_3 + k_5} \Big[\left(k_3 - k_5\right) \left(\Delta^2 + \Omega^2 \cos\! \left(\Omega_R T_{p2}\right)\right) \Big. \\
	& \quad\Big. - 2 \Omega^2 \left(k_1 - k_2\right) e^{(\gamma_1 + \gamma_2 - \gamma_d)\frac{T}{2}}\Big].
\label{eq:ramseyvar1}
\end{aligned}
\end{equation}

If an equal splitting is desired at an arbitrary detuning, the splitter pulse duration $T_{p1}$ can be obtained from the lossless model \cite{Ivannikov2013thesis} by solving the equation $P_z(t) = 0$:
\begin{equation}
\begin{aligned}
\rho_{11} &= \frac{1}{2 \Omega_R^2}\left(\Delta^2+\Omega_R^2+\Omega^2\cos\left(\Omega_R t\right)\right), \\
\rho_{22} &= \frac{\Omega^2}{2 \Omega_R^2}\left(1-\cos\left(\Omega_R t\right)\right).
\end{aligned}
\label{eq:losslessrabi}
\end{equation}
The first pulse duration is then
\begin{equation}
t_{\pi/2} = T_{p1} = \left.\arccos\!\left(-\frac{\Delta^2}{\Omega^2}\right)\right/\Omega_R^2,
\label{eq:tp1pi2}
\end{equation}
where the sequence can be closed by a $\pi/2$-pulse defining the duration $T_{p2} = \pi/(2\Omega_R)$. The $T_{p1}$ is limited by the detuning that is required to be not larger than the resonant Rabi frequency: $\Omega \geq \Delta$. If this condition is not satisfied, the equation $P_z(t) = 0$ gives an unphysical imaginary result. The maximal possible off-resonant $\pi/2$-pulse duration that provides equal population splitting is then found to be $t_{\pi/2} = \pi \left/\left(\Omega\sqrt{2}\right)\right.$.

A direct application of the $T_{p2}$ variation is to model the experimental imperfections, associated with the second pulse. The $T_{p1}$ variation is of a more subtle character: The splitter allows us to initiate free evolution with unequal populations that may evolve into \textit{equal} populations. For this to happen, the state with the higher loss rate needs to be loaded more at the beginning of the evolution. The point where the unequally split populations equalize gives the maximal normalized population value. By chasing the optimal value of $T$ by accordingly correcting the splitter duration $T_{p1}$ one can attain a perpetually maximal contrast of $P_z(T)$. This is of benefit for data fitting since the envelope function becomes constant. Of course, this method does not affect the signal-to-noise ratio defined by the fundamental limit, the Heisenberg uncertainty.

\subsection{One-body loss asymmetry cancellation}
\label{sec:asymcancel}
% 13-01-27. Rabi pedestal in Ramsey spectrum.nb
% 13-01-25. Atom clock, Ramsey resonance.nb
\begin{figure*}[t]
\includegraphics[width=1.0\textwidth]{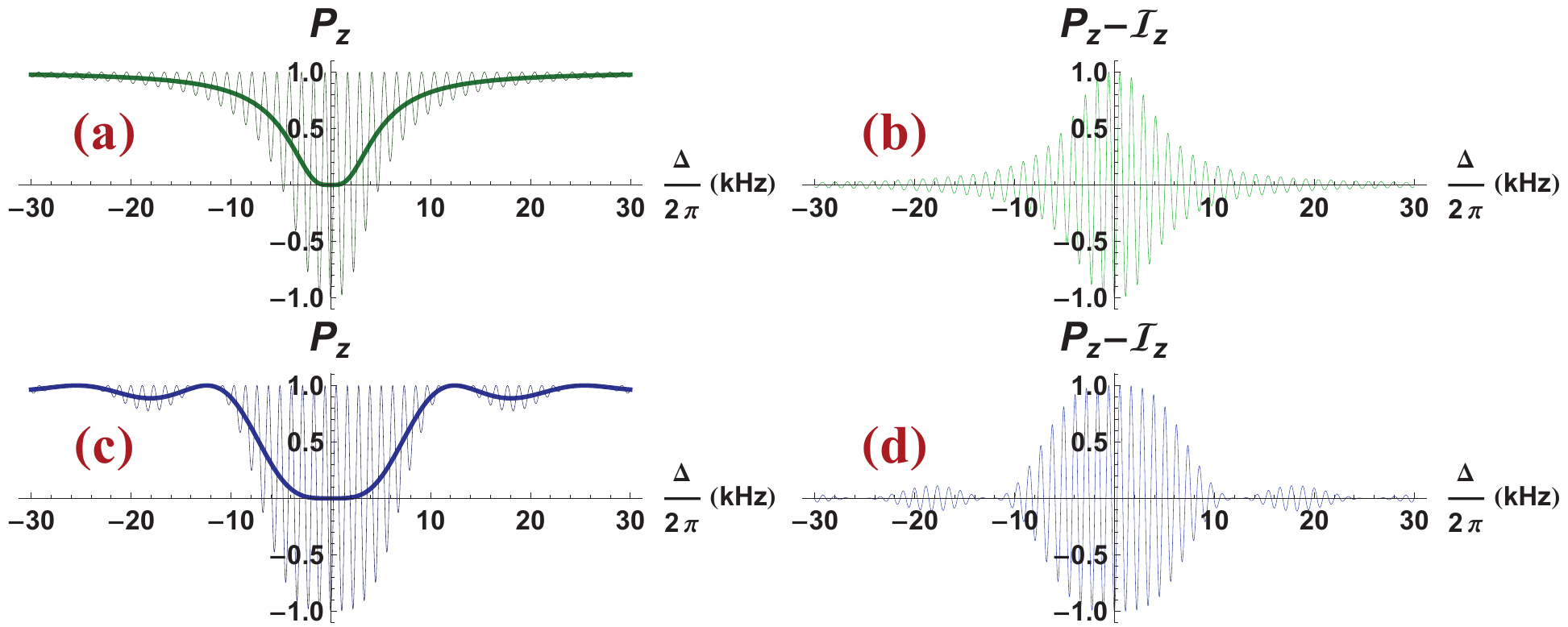}
\caption[Ramsey spectra of the two analytical models]{(Color available online) Ramsey spectra. The following values are used: $\Omega = 2\pi\times (510\text{~Hz})$, $T = 5$~ms. The measurable $P_z(\Delta,T)$ with the Ramsey condition $\varphi = \Delta$ (the one-body model of Eqs.~(\ref{eq:ramseyFormulae4})) is plotted in \textbf{(a)}. Losses are neglected. The Rabi pedestal in \textbf{(a)} has a narrow line shape at $T=0$ and has a single-peak coming from Eq.~(\ref{eq:ramseyvar1}) where the pulses are assumed to be $\pi/2$ for all $\Delta$. Whereas in \textbf{(c)}, corresponding to the model of Eqs.~(\ref{eq:ramseyFormulae4}), the pulses are $\pi/2$ only at resonance; at $\Delta\neq 0$ they become non-$\pi/2$-pulses distorting the spectrum. \textbf{(b)} and \textbf{(d)} are differences $P_z - \mathcal{I}_z$ of \textbf{(a)} and \textbf{(c)}, correspondingly.}
\label{fig:clockresonance}
\end{figure*}%

The visibility in long Ramsey experiments decreases due to loss asymmetry as one of the dominant factors \cite{Ivannikov2013thesis}. A look at the population decays in Figs.~\ref{fig:asymcompens}(a,b) suggests that if the populations start from unequal values $N_2(T=0) > N_1(T=0)$, then $N_1(T)$ and $N_2(T)$ cross. $P_z$ has a maximum at this point; i.e., the loss asymmetry is cancelled. One can tailor a sequence with variable splitter $\pi/2$-pulse duration to obtain a Ramsey fringe that gives unitary visibility at a desired location. To derive the expression for the optimal first-pulse duration, the density operator $\rho(t)$ is propagated until the end of free evolution, before the second $\pi/2$-pulse, where the populations are
\begin{equation}
\begin{aligned}
\rho_{11}(t) &= \frac{1}{2 \Omega_R^2} e^{-\gamma_1 t} \left(\Delta^2+\Omega_R^2+\Omega^2 \cos\left(\Omega_R T_{p1}\right)\right), \\
\rho_{22}(t) &= \frac{1}{2 \Omega_R^2} e^{-\gamma_2 t} \Omega^2 \big(1 - \cos\left(\Omega_R T_{p1}\right)\big).
\end{aligned}
\label{eq:lossprob}%
\end{equation}
Then the crossing of the populations is found by solving equation $\rho_{11}(t) = \rho_{22}(t)$ with respect to time $t$ and discarding irrelevant solutions. The solution gives time where the maximum visibility of $P_z$ occurs as a function of $T_{p1}$; we label this time $T_\text{optimal}$ further on:
\begin{equation}
T_\text{optimal} = \frac{1}{\gamma_1-\gamma_2}\ln \Bigg(\frac{\Omega^2_R}{\Omega^2} \csc^{2}\left(\frac{\Omega_R T_{p1}}{2}\right)-1\Bigg).
\label{eq:optimalt1}
\end{equation}

In Fig.~\ref{fig:asymcompens} the effect of loss compensation is shown with a set of test parameters. Figure~\ref{fig:asymcompens}(a) shows the dynamics of the freely evolving populations following the application of the standard $\pi/2$ splitting pulse. The populations are plotted before the arrival of the detecting $\pi/2$-pulse. In contrast, Fig.~\ref{fig:asymcompens}(b) shows how the splitter can affect the populations and lead to their balance at an arbitrary time $T$. Figures.~\ref{fig:asymcompens}(c,d) show the populations and measurable $P_z$ after the full interferometric sequence with a non-zero $\varphi$ producing a fringe. In accordance with the expectations, Fig.~\ref{fig:asymcompens}(d) indicates an extremum in the visibility at $T = T_\text{optimal}$ defined by Eq.~(\ref{eq:optimalt1}).

\section{Rabi pedestals \& Ramsey spectra}
\label{sec:atomclockop}

As in the case of the Rabi model of Eqs.~(\ref{eq:losslessrabi}), the Ramsey spectrum (Fig.~\ref{fig:clockresonance}, lossless, $\varphi = \Delta$) has a comb of resonances at around $\Delta = 0$ that narrow down with increasing evolution time $T$. At resonance the visibility is highest and the slope is steepest, which is ultimately converted to the best interferometer accuracy.

For the measurables $P_1$, $P_2$, and $P_z$ the corresponding Rabi pedestal \cite{Sullivan2001} functions $\mathcal{I}_1$, $\mathcal{I}_2$, and $\mathcal{I}_z$ with always resonant $\pi/2$-pulses given by $T_{p1}=T_{p2}=\pi/2/\Omega_R$, forming the baseline for the Ramsey oscillations, are $\{\frac{1}{2}+\frac{g}{2},\frac{1}{2}-\frac{g}{2},g\}$, where $g=\Delta^4/\Omega_R^4$. It is remarkable that the pedestals, and, consequently, the oscillation envelopes, are more flat at around $\Delta=0$, than the Lorentzians of the Rabi spectra \cite{Ivannikov2013thesis}. More general pedestal functions are obtained by averaging the measurables from the variable-pulse model:
\begin{equation}
\begin{aligned}
2 \mathcal{I}_1 \Omega_R^4&=(2+a+b) \Delta^2 \Omega^2+(1+a b) \Omega^4+2 \Delta^4,\\
2 \mathcal{I}_2 \Omega_R^4&=\left(\Omega^2+\Delta^2\left(2-b\right)-a \left(\Delta^2+b \Omega^2\right)\right)\Omega^2,\\
\mathcal{I}_z \Omega_R^4&=\left(a\Omega^2+\Delta^2\right) \left(b\Omega^2+\Delta^2\right),
\end{aligned}%
\label{eq:pedestalsstd}%
\end{equation}%
with $a = \cos \left(\Omega_R T_{p1}\right)$ and $b = \cos \left(\Omega_R T_{p2}\right)$. The averages of the standard Ramsey pulses can be modelled by setting $T_{p1}=T_{p2}=\pi/2/\Omega$ to Eqs.~(\ref{eq:pedestalsstd}). These baseline functions only depend on the pulse parameters $\Omega$, $\Delta$, $T_{p1}$, and $T_{p2}$. Hence, they isolate Ramsey-interference and Rabi-pulse-related contributions. Differences $\{P_1-\mathcal{I}_1, P_2-\mathcal{I}_2, P_z-\mathcal{I}_z\}$ contain only Ramsey-related interference patterns (Figs.~\ref{fig:clockresonance}(b,d)).

In the model of Eqs.~(\ref{eq:ramseyFormulae4}) the $\pi/2$-pulses split the populations of the two states $50$:$50$, even off resonance with $\Omega \neq \Omega_R$. This is different from the standard $\pi/2$-pulse whose duration is adjusted while at resonance and kept constant when the detuning $\Delta$ is changed. Such pulses with the $\Delta$-dependent duration produce a Rabi pedestal with a single broad peak as shown in Fig.~\ref{fig:clockresonance}(a). In the Ramsey approach the $\pi/2$-pulse durations are kept constant, i.e., $T_{p1} = T_{p2} = \pi/2\left/\Omega\right.$ in Eq.~(\ref{eq:ramseyvar1}). In this case the pulses split the populations into halves at resonance, but provide an unequal splitting away from resonance. The corresponding Ramsey spectrum has an infinite sequence of maxima in the envelope function in Fig.~\ref{fig:clockresonance}(c). Losses in Fig.~\ref{fig:clockresonance} are neglected for they are system specific; however, in a more realistic model the effects of the Maxwell-Boltzmann velocity distribution, atomic motion and miscellaneous inhomogenieties \cite{Rosenbusch2009} may suppress or distort the Ramsey features away from $\Delta=0$, resulting in a spectrally more localized envelope \cite{Ramsey1956book,Xu1999}.

These two models (Fig.~\ref{fig:clockresonance}) have different assumptions about how the $\pi/2$-pulse duration is defined in experiment. Typically in applications, near resonant operation is desirable to benefit from high visibility; hence in the present discussion we neglect the contrast loss away from $\Delta = 0$, and the envelope shift away from $\Delta=0$ and we limit ourselves to the dominant effect of $\Omega_R$ solely forming the broad spectral envelope in Fig.~\ref{fig:clockresonance}. Individual features of the spectral comb are also affected by the collisional shift \cite{Harber2002} and the effects of the trap \cite{Rosenbusch2009}; e.g., phase difference acquired during the evolution, if any, shifts the interference pattern. In clocks this effect is undesirable, but it is routinely used in sensing applications.

\section{Conclusions}

In this work we presented a set of Ramsey-type models in which Ramsey interference and Rabi-pulse-related pedestal were separated, and one-body models were generalized into a variable-duration pulse model with the detunings separately defined for the periods of pulse coupling and free evolution.

The presented models are of general interest; they can be employed when many-body physics is negligible. A many-body model would be described by a system of nonlinear differential equations difficult to solve in an exact analytical form \cite{Ivannikov2014}. The one-body model is parametrized with the detuning and with variable pulse durations to model realistic systems where pulses may be generated with imperfections or be distorted by the medium. Such a flexible model allows to find an optimal splitter-pulse duration that cancels the effect of unequal one-body losses on the $P_z$ visibility \cite{Ivannikov2013thesis}. It turns out that the expression for the optimal splitter pulse duration has a simple analytical form. The cancellation strategy can be extended to the many-body case; however, a greater number of decay channels would ensue a more complex analysis.

Equation~(\ref{eq:optimalt1}) is valid for one-body-decay limited systems. The many-body counterpart of Eqs.~(\ref{eq:masterexplicit}) can also be obtained \cite{Ivannikov2014}; in practice this implies numerical integration to search the corresponding $T_{\text{optimal}}$. The attained effect of constant $P_z(\phi,T)$ visibility can facilitate, e.g., atom-clock stability analysis. It should be noted that this technique does not improve the signal-to-noise ratio. The approach is equally valid for time- and phase-domain Ramsey experiments \cite{Ivannikov2013thesis}.

\begin{acknowledgments}
The author thanks Andrei Sidorov and Peter Hannaford for stimulating discussions.
\end{acknowledgments}

\appendix
\section{Off-resonant spin-echo interferometry with one-body losses}
\label{sec:measurables}

% 12-10-04. Thesis calculations 7 (spin echo).nb
% 12-10-01. Thesis calculations 6 (Ramsey) v2 - fixed.nb
\begin{figure*}[pt]
\includegraphics[width=1.0\textwidth]{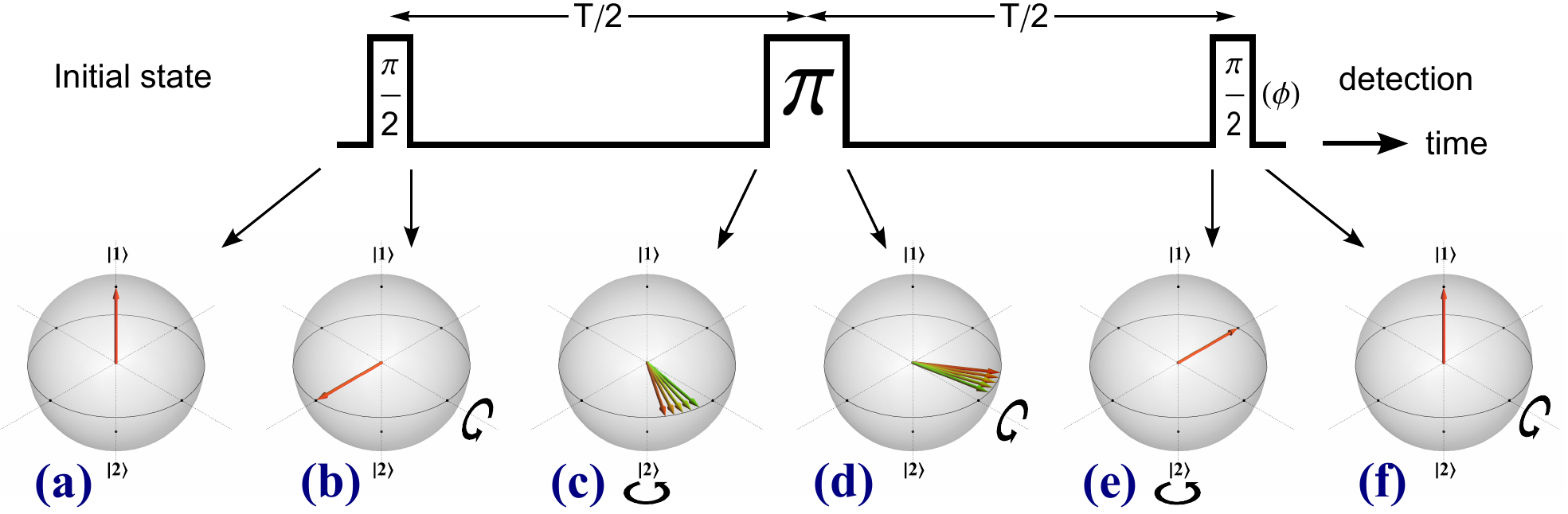}
\caption[Bloch vector evolution in spin-echo interferometry]{(Color available online) Spin-echo interferometry in the Bloch vector representation. $\phi$ is the phase of the second $\pi/2$-pulse. As in Ramsey interferometry (Fig.~\ref{fig:blochramsey}), $\phi$ modulation can be used at a fixed $T$ to record a $P_z$ fringe. Insets \textbf{(a)}, \textbf{(b)}, \textbf{(c)} show dynamics identical to that of Figs.~\ref{fig:blochramsey}(a,b,c). In \textbf{(d)} the spin-echo pulse implements phase reversal, and the ensemble starts rephasing. \textbf{(e)} After the second $\pi/2$-pulse the Bloch vectors are refocused. \textbf{(f)} The ensemble spin spread is nullified.}
\label{fig:blochspinecho}
\end{figure*}

In spin-echo interferometry (Fig.~\ref{fig:ramseyvar}(b)) $\mathbf{B}$ also undergoes rotations by $\mathbf{\Omega}$ (Fig.~\ref{fig:blochspinecho}). In the middle of the sequence a phase-reversal $\pi$-pulse flips the dephasing direction, and the pseudo-spins start rephasing and refocusing by the second $\pi/2$-pulse. The solution of Eq.~(\ref{eq:master}) for an off-resonant spin-echo sequence with losses during free evolution is parametrized by the total sequence duration $T$, and the cumulative Ramsey dephasing rates in the two arms of the interferometer $\varphi_1$ and $\varphi_2$ that allow one to include miscellaneous physical mechanisms displacing the energy levels.

It is convenient to isolate interference from decay terms
\begin{equation}
\begin{aligned}
 a_0 &\!=\! e^{-\gamma_3 T}\!,\,
 a_1 \!=\! e^{-\gamma_1 T}\!,\,
 a_2 \!=\! e^{-\frac{\gamma_1+\gamma_3}{2} T}\!,\,
 a_3 \!=\! e^{-\gamma_2 T}\!,\\
 a_4 &\!=\! e^{-\frac{\gamma_2+\gamma_3}{2}T},\,
 a_5 \!=\! e^{-\frac{\gamma_1+\gamma_2}{2} T}\!,\,
 a_6 \!=\! \cos\!\left(\frac{\varphi_1}{2}T\right)\!,\,\\
 a_7 &\!=\! \cos\!\left(\frac{\varphi_2}{2}T\right)\!,\,
 a_8 \!=\! \sin\!\left(\frac{\varphi_1}{2}T\right)\!,\,
 a_9 \!=\! \sin\!\left(\frac{\varphi_2}{2}T\right)\!,\,\\
 b_0 &\!=\! e^{\left(\gamma_1 + \frac{\gamma_3}{2}\right) T}\!,\,
 b_1 \!=\! e^{\frac{1}{2} (\gamma_1+\gamma_2+\gamma_3) T}\!,\,
 b_2\! =\! e^{\left(\gamma_1+\frac{\gamma_2}{2}\right) T}\!,\,\\
 b_3 &\!=\! e^{\left(\gamma_2 + \frac{\gamma_3}{2}\right)T}\!,\,
 b_4 \!=\! e^{\left(\frac{\gamma_1}{2}+\gamma_2\right)T}\!,\,
 b_5 \!=\! e^{\left(\gamma_1+\gamma_2-\frac{\gamma_3}{2}\right) T}\!,\\
k_0 &\!=\! \Delta^2 \Omega^2 b_0 \Omega_R^3+\Delta^4 b_3 \Omega_R^3-2 \Delta^2 \Omega^2 b_2 a_6 \Omega_R^3 \\
	& \quad +2 \Omega^2 b_1 \Omega_R^5\! + \!2 \Delta^2 \Omega^2 b_4 a_6 \Omega_R^3 \! - \!2 \Delta \Omega^2 b_2 a_8 \Omega_R^4 \\
	& \quad +2 \Delta  \Omega^2 b_4 a_8 \Omega_R^4\!\! +\!\Delta^2 b_3 \Omega_R^5,
\end{aligned}
\label{eq:definitions}
\end{equation}
from $\rho_{11}$, $\rho_{22}$, and $P_z$:
\begin{subequations}
\begin{align}
\!\!\!\!\rho_{11}\cdot & 4 \Omega_R^7 = \Delta^6 a_1 \Omega_R -2 \Delta^3 \Omega^4 a_4 a_8-2 \Delta \Omega^6 a_4 a_8\nonumber\\
		&\!\!\!\!\!\!\!\!-4 \Delta^5 \Omega^2 a_0 a_7 a_8-4 \Delta^3 \Omega^4 a_0 a_7 a_8+\Delta^2 \Omega^4 a_3 \Omega_R\nonumber \\
		&\!\!\!\!\!\!\!\! +2 \Delta^2 \Omega^4 a_5 \Omega_R+2 \Delta^4 \Omega^2 a_2 a_6 \Omega_R-2 \Delta^2 \Omega^4 a_4 a_6 \Omega_R \nonumber \\
		&\!\!\!\!\!\!\!\! +2 \Delta^4 \Omega^2 a_2 a_7 \Omega_R-2 \Delta^2 \Omega^4 a_4 a_7 \Omega_R+\Delta^2 a_1 \Omega_R^5\nonumber \\
		&\!\!\!\!\!\!\!\! +2 \Omega^6 a_0 a_6 a_7 \Omega_R+2 \Delta^3 \Omega^2 a_2 a_8 \Omega_R^2+2 \Delta^3 \Omega^2 a_2 a_9 \Omega_R^2\nonumber \\
		&\!\!\!\!\!\!\!\! -2 \Delta \Omega^4 a_4 a_9 \Omega_R^2-4 \Delta^3 \Omega^2 a_0 a_6 a_9 \Omega_R^2 \nonumber \\
		&\!\!\!\!\!\!\!\! +2 \Delta^4 a_1 \Omega_R^3+2 \Omega^4 a_5 \Omega_R^3+2 \Delta^2 \Omega^2 a_2 a_6 \Omega_R^3\nonumber \\
		&\!\!\!\!\!\!\!\! +2 \Delta^2 \Omega^2 a_2 a_7 \Omega_R^3+2 \Omega^4 a_0 a_8 a_9 \Omega_R^3+2 \Delta \Omega^2 a_2 a_8 \Omega_R^4\nonumber \\
		&\!\!\!\!\!\!\!\! +2 \Delta \Omega^2 a_2 a_9 \Omega_R^4 +6 \Delta^2 \Omega^4 a_0 a_6 a_7 \Omega_R,\\[2mm]
\!\!\!\!\rho_{22}\cdot & 4 \Omega_R^7 = \Delta^4 \Omega^2 a_1 \Omega_R+\Delta^4 \Omega^2 a_3 \Omega_R-4 \Delta^5 \Omega^2 a_4 a_8\nonumber \\
		&\!\!\!\!\!\!\!\! -6 \Delta^3 \Omega^4 a_4 a_8-2 \Delta  \Omega^6 a_4 a_8 +4 \Delta^3 \Omega^4 a_0 a_7 a_8\nonumber \\
		&\!\!\!\!\!\!\!\! +4 \Delta^5 \Omega^2 a_0 a_7 a_8+4 \Delta^4 \Omega^2 a_5 \Omega_R+4 \Delta^2 \Omega^4 a_5 \Omega_R\nonumber \\
		&\!\!\!\!\!\!\!\! -2 \Delta  \Omega^2 a_2 a_9 \Omega_R^4 +2 \Omega^6 a_5 \Omega_R+2 \Delta^2 \Omega^4 a_2 a_6 \Omega_R\nonumber \\
		&\!\!\!\!\!\!\!\! -4 \Delta^4 \Omega^2 a_4 a_6 \Omega_R-2 \Delta^2 \Omega^4 a_4 a_6 \Omega_R -2 \Delta^4 \Omega^2 a_2 a_7 \Omega_R\nonumber \\
		&\!\!\!\!\!\!\!\! +2 \Delta^2 \Omega^4 a_4 a_7 \Omega_R-6 \Delta^2 \Omega^4 a_0 a_6 a_7 \Omega_R-2 \Omega^6 a_0 a_6 a_7 \Omega_R \nonumber \\
		&\!\!\!\!\!\!\!\! +2 \Delta  \Omega^4 a_2 a_8 \Omega_R^2-2 \Delta^3 \Omega^2 a_2 a_9 \Omega_R^2+2 \Delta  \Omega^4 a_4 a_9 \Omega_R^2\nonumber \\
		&\!\!\!\!\!\!\!\! +4 \Delta^3 \Omega^2 a_0 a_6 a_9 \Omega_R^2+\Delta^2 \Omega^2 a_1 \Omega_R^3+\Delta^2 \Omega^2 a_3 \Omega_R^3\nonumber \\
		&\!\!\!\!\!\!\!\! -2 \Delta^2 \Omega^2 a_2 a_7 \Omega_R^3-2 \Omega^4 a_0 a_8 a_9 \Omega_R^3, \\[2mm]
P_z\cdot & k_0 = 2 \Delta^5 \Omega^2 b_2 a_8-4 \Delta^3 \Omega^2 b_5 a_6 a_9 \Omega_R^2+2 \Delta  \Omega^2 b_4 a_9 \Omega_R^4\nonumber \\
		&\!\!\!\!\!\!\!\! +2 \Delta^3 \Omega^4 b_4 a_8-4 \Delta^5 \Omega^2 b_5 a_7 a_8 -4 \Delta^3 \Omega^4 b_5 a_7 a_8\nonumber \\
		&\!\!\!\!\!\!\!\! -\Delta^4 \Omega^2 b_0 \Omega_R-2 \Delta^4 \Omega^2 b_1 \Omega_R+2 \Delta^6 b_3 \Omega_R+\Delta^4 \Omega^2 b_3 \Omega_R \nonumber \\
		&\!\!\!\!\!\!\!\! +2 \Delta^4 \Omega^2 b_2 a_6 \Omega_R+2 \Delta^4 \Omega^2 b_4 a_6 \Omega_R-2 \Delta^2 \Omega^4 b_2 a_7 \Omega_R\nonumber \\
		&\!\!\!\!\!\!\!\! +4 \Delta^4 \Omega^2 b_4 a_7 \Omega_R+2 \Delta^2 \Omega^4 b_4 a_7 \Omega_R+6 \Delta^2 \Omega^4 b_5 a_6 a_7 \Omega_R\nonumber \\
		&\!\!\!\!\!\!\!\! +2 \Omega^6 b_5 a_6 a_7 \Omega_R-2 \Delta  \Omega^4 b_2 a_9 \Omega_R^2+2 \Delta^3 \Omega^2 b_4 a_9 \Omega_R^2\nonumber \\
		&\!\!\!\!\!\!\!\! +2 \Delta^3 \Omega^4 b_2 a_8+2 \Omega^4 b_5 a_8 a_9 \Omega_R^3+2 \Delta^5 \Omega^2 b_4 a_8.
\end{align}
\label{eq:spinechoFormulae4}%
\end{subequations}
Provided the detuning is zero, which is physically justified in the case of $\Delta \ll \Omega$, Eqs.~(\ref{eq:spinechoFormulae4}) become
\begin{subequations}
\begin{align}
\rho_{11} &= \frac{1}{2} e^{-\gamma_3 T} \left[e^{\frac{\gamma_d}{2}T}+\cos\! \left(\frac{\varphi_1-\varphi_2}{2} T \right)\right],\\
\rho_{22} &= \frac{1}{2} e^{-\gamma_3 T} \left[e^{\frac{\gamma_d}{2}T}-\cos\! \left(\frac{\varphi_1-\varphi_2}{2} T \right)\right],\\
P_z &= e^{-\frac{1}{2}\gamma_d T}\cos\!\left(\frac{\varphi_1-\varphi_2}{2}T\right).
\end{align}
\label{eq:spinechoFormulae2}%
\end{subequations}
It follows from Eqs.~(\ref{eq:spinechoFormulae2}) that the phase acquired in the first arm of the spin-echo interferometer $\left.\varphi_1 T\right/2$ is completely eliminated by the same value of the phase in the second arm $\left.\varphi_2 T\right/2$, which is the expected behavior.

%\bibliographystyle{plain}
%\bibliography{thesis}


\end{document}